\documentclass[a4paper,fleqn]{cas-dc}

\usepackage[authoryear,longnamesfirst]{natbib}
\usepackage{comment}
\usepackage{multirow}
\usepackage{longtable}
\usepackage{graphicx}
\usepackage{makecell}
\usepackage{pifont}
\usepackage{hyperref}
\usepackage{enumitem}
\usepackage{afterpage}
\usepackage{subfig}

\makeatletter
\newcommand{\aftertwo}[1]{\afterpage{\if@firstcolumn #1
  \else\afterpage{#1}\fi}}
\makeatother
\def\tsc#1{\csdef{#1}{\textsc{\lowercase{#1}}\xspace}}
\tsc{WGM}
\tsc{QE}
\tsc{EP}
\tsc{PMS}
\tsc{BEC}
\tsc{DE}

\begin{document}
\let\WriteBookmarks\relax
\def\floatpagepagefraction{1}
\def\textpagefraction{.001}

\shorttitle{False Data Injection Threats in Active Distribution Systems: A Comprehensive Survey}

\shortauthors{M. A. Husnoo et~al.}

\title [mode = title]{False Data Injection Threats in Active Distribution Systems: A Comprehensive Survey}                      
\tnotemark[1]

\tnotetext[1]{This work is the results of the research project funded by the Faculty of Science, Engineering and Built Environment (SEBE), Deakin University under the scheme 'Mini ARC Analogue Program (MAAP)'.}

\author[1]{Muhammad Akbar Husnoo}[orcid=0000-0001-7908-8807]
\ead{mahusnoo@deakin.edu.au}

\address[1]{Deakin University, Geelong, Australia - Centre for Cyber Security Research and Innovation (CSRI)}

\author[1]{Adnan Anwar}[orcid=0000-0003-3916-1381]
\cormark[1]
\fnmark[2]
\ead{adnan.anwar@ deakin.edu.au}

\author[2]{Nasser Hosseinzadeh}
\ead{nasser.hosseinzadeh@deakin.edu.au}

\address[2]{Deakin University, Geelong, Australia - Centre for Smart Power and Energy Research (CSPER)}

\author[2]{Shama Naz Islam}[orcid=0000-0002-2354-7960]
\ead{shama.i@deakin.edu.au}

\author[3]{Abdun Naser Mahmood}[orcid=0000-0001-7769-3384]
\ead{A.Mahmood@latrobe.edu.au}

\address[3]{Department of Computer Science \& IT, Latrobe University, Bundoora, VIC 3086, Australia}

\author[1]{Robin Doss}
\ead{robin.doss@deakin.edu.au}

\cormark[1]

\begin{abstract}
With the proliferation of smart devices and revolutions in communications, electrical distribution systems are gradually shifting from passive, manually-operated and inflexible ones, to a massively interconnected cyber-physical smart grid to address the energy challenges of the future. However, the integration of several cutting-edge technologies has introduced many security and privacy vulnerabilities due to the large-scale complexity and resource limitations of deployments. Recent research trends have shown that False Data Injection (FDI) attacks are becoming one of the most malicious cyber threats within the entire smart grid paradigm. Therefore, this paper presents a comprehensive survey of the recent advances in FDI attacks within active distribution systems and proposes a taxonomy to classify the FDI threats with respect to smart grid targets. The related studies are contrasted and summarized in terms of the attack methodologies and implications on the electrical power distribution networks. Finally, we identify some research gaps and recommend a number of future research directions to guide and motivate prospective researchers.
\end{abstract}



\begin{keywords}
False Data Injection Attack \sep Distribution System \sep Smart Meter \sep Advanced Metering Infrastructure \sep AMI \sep Smart Grid
\end{keywords}

\maketitle

\section{Introduction}
\label{sec:Introduction}
In this new global economy, adequacy of uninterrupted power supply to end-users has now become one of the main priorities of the critical energy infrastructure of several nations. In the past few years, traditional manually-operated distribution systems have shifted to smart distribution systems to cope up with the increased power consumption demands and operational reliability \cite{Pahwa_2015}. Recently, the growing trend of integrating distributed renewable energy sources into power systems have also shifted focus from passive distribution systems into Active Distribution Systems (ADSs) in response to environmental concerns, power sustainability and energy market economics \cite{Radwan_Zaki}. 

In line with the IEEE Grid Vision 2050, the main goal of a smart grid is to enable efficient and reliable bi-directional communication through control and automation processes applied through the different components of a power grid \cite{6577603}. This vision is being achieved by converting the static grid into intelligent cyber-physical systems through the integration of information and communication technologies. Modern technologies including Internet of Things (IoT), and cloud computing are considered to be the foundations of ADSs, which take full advantage of such cutting-edge technologies to proactively coordinate the renewable energy generation, energy storage and other distributed units in view of achieving safe and economical operation of smart grids \cite{Yang_2019}. 

\begin{figure*}
  \centering
  \includegraphics[width=16cm]{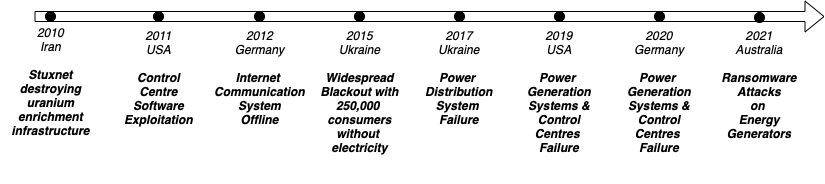}
  \caption{Timeline of recent power grid attacks (A chronological timeline of the major power grid attacks that have occurred throughout the globe starting from 2010 up until 2021). \label{powerattacks}}
\end{figure*}

There has been a massive shift of research focus from transmission systems to distribution systems as the latter is highly influenced by socio-economic and environmental parameters given its close proximity to end-users \cite{8887286}. However, the rush to integrate of a wide variety of technologies and components with distribution systems has neglected the security aspects of ADSs. Furthermore, the bi-directional communication within ADSs brings additional security and privacy challenges due to the large-scale complexity and resource limitations of deployments \cite{Jokar_Arianpoo_Leung_2016}. Coupled with limited cybersecurity research done on distribution networks, there are growing concerns over massive threats potentially impacting its integrity, reliability and stability. From a cyber-physical perspective, the success of distribution systems is heavily reliant on its physical components which is the power network infrastructure, and its cyber components which includes information sensing, data analytics, etc \cite{https://doi.org/10.48550/arxiv.0912.5233}. As earlier mentioned, attacks on transmission systems have been well-studied in the past \cite{5622046}. An interesting attack proposed by \cite{7636987} that considers physical tampering by disconnecting some power lines as well as cyber manipulation by blocking the information flow was used to disrupt the grid operability. Indeed, such attacks concepts may eventually be extended to distribution systems \cite{6848211}. One such notorious cyber-physical attack on distribution networks happened on 23\textsuperscript{rd} December 2015 at Kiev, Ukraine, where perpetrators gained unauthorized access to the Supervisory Control and Data Acquisition (SCADA) system and tampered with circuit breakers which affected more than 225,000 customers for several hours \cite{7752958}. Figure \ref{powerattacks} shows a timeline of the well-known attacks that have occurred on power grid systems throughout the recent years.

\subsection{Motivation \& Scope}

With the integration of IoT-based devices into power grid systems, cybersecurity is now a prominent aspect of smart grids due to the wide array of security attacks. Attacks and countermeasures within the transmission network of power grids have been well studied in several previous literature e.g. works conducted by the authors in \cite{Jokar_Arianpoo_Leung_2016, 6129371, 7579185}. With the rise in the number of cyber-related incidents on smart grids such as the Ukraine 2015 outage \cite{7752958}, it is obvious that such attacks can have devastating consequences on ADSs. While current research is mainly focused on the active applications of cutting-edge technologies and the development of enhanced communications methods within distribution networks, the security risks introduced are seldom considered from the perspective of adversaries. At present, very little work has been done in relation to discovering the vulnerabilities of distribution systems which leads to urgent calls in ensuring the resilience of ADSs to different types of emerging threats and zero-day attacks.

In recent years, False Data Injection threats (FDI) have surfaced as one of the most critical threats faced by distribution systems whereby an adversary deliberately introduces corrupted noisy information into sensor measurements while being as stealthy as possible. As earlier mentioned, FDI threats in transmission systems have been studied and some countermeasures have been proposed. Indeed, the security aspects of ADSs have been previously neglected. Since scholars and researchers are now shifting their focus on improving the emerging security and privacy aspects of ADSs, it is critical to have an understanding of the characteristics of distribution systems which emphasise the needs and motivation of ongoing research on FDI threats and countermeasures:
\begin{enumerate}[wide, labelwidth=!, labelindent=0pt]
    \item Lack of proper security mechanisms: Power distribution networks consist of several field devices including power transformers, voltage regulators, remote terminal units and so on. Such types of field devices lack proper cyber and physical security mechanisms \cite{Otuoze_Mustafa_Larik_2018}. Indeed, without strong security measures, adversaries can easily gain access and construct data integrity threats, specifically FDI attacks.
    
    \item Communication Protocols: In a distribution system, Supervisory Control And Data Acquisition (SCADA) and/or Industrial IoT (IIoT) protocols such as ZIGBEE, Modbus, etc. are widely used for substation communication or communication between substation and field devices \cite{Anwar_Mahmood_Tari_Kalam_2022}. These protocols are vulnerable to security threats. IEEE, IEC and other standard organisations who have been involved in the development of protocols are actively researching on increasing the resiliency of these protocols against cyber threats. However, widely used protocols, e.g. Modbus or Zigbee, still lack proper security mechanisms. As a result, messages sent without proper strong encryption or authentication mechanisms can be exposed to FDI threats \cite{McLaughlin_Friedberg_Kang_Maynard_Sezer_McWilliams_2015}.
    
    \item Distribution System Characteristics: As earlier highlighted, power distribution systems have different characteristics as compared to transmission systems such as high R/X ratio, etc \cite{8890222}. Therefore, the formulation of FDI threats on transmission systems will work differently in a distribution system. The assumption of DC power flow formulation is not valid for distribution systems. Moreover, the AC formulation for distribution systems need to consider multi-phase power flow models and distribution state estimations \cite{9444771}. Hence, it is very important to investigate the both the attack construction and the detection of FDI threats on distribution systems.
    
    \item Heterogeneity: The wide array of IoT devices in modern power distribution system e.g. smart homes, SCADA, field devices, sub-station devices, etc. along with the different communication protocols and different stakeholders within a single network increases the complexity of the network. For maintenance and monitoring purposes, third party companies need to connect to the proprietary networks through internet, which also introduces more vulnerabilities \cite{6257527, Renugadevi_Saravanan_Naga}. Eventually, with increasing complexity, the window of opportunities for FDI attacks increases. 
\end{enumerate}

Therefore, the primary objective of this manuscript is to provide an early systematic literature review and insight into the FDI threats within active distribution systems to motivate future researchers into exploring some of the emerging areas within active ADSs.

\subsection{Contributions}

This manuscript covers a comprehensive survey of the existing publications and reference materials on cyber threats identified within the various domains of the smart grid distribution infrastructure. Our primary objective is to systematically and thoroughly analyze recent literature within the last decade, and assess and contrast each proposed attack methodology. In particular, the main contributions of our article are listed as follows: 
\begin{enumerate}[wide, labelwidth=!, labelindent=0pt]
    \item We identify the essential cybersecurity goals of active distribution systems and provide some theoretical overview of stealthy FDI attacks.
    
    \item Following a comprehensive review of the relevant existing literature, we highlight their contributions and identify the gaps as addressed by our survey. A detailed comparison of previous works against ours can be found in Table \ref{tab:comparison}.
    
    \item We develop and propose a detailed taxonomy of FDI attacks with respect to attack targets in active distribution systems as shown in Section \ref{sec:proposedtaxonomy}.
    
    \item We analyze the various state-of-the-art FDI threat modeling proposed by several independent studies, critically evaluate the approaches and summarize the results. 
    
    \item Lastly, we discuss some main research gaps in the existing FDI attack methodologies and provide some technical recommendations for future research directions within the related topic.

\end{enumerate}

We strongly believe that an early systematic review of independently developed research will enable future researchers to get a clear picture of the neglected security aspects of AMI-based ADSs, thus contributing towards more resilient distribution systems of the future.

\subsection{Paper Structure}

\begin{table}[]
    \centering
    \begin{tabular}{|p{2cm}|p{5cm}|}
        \hline
         \textbf{Terminologies} &  \textbf{Definitions}\\ \hline
         
         Active Distribution System (ADS) & Distribution system with operation and control capabilities coupled with distributed decentralized energy resources. \\ \hline
         
         Advanced Metering Infrastructure (AMI) & Network of smart meters and data management systems that allows bi-directional communication between the utility providers and the consumers. \\ \hline
         
         Control Center & Consists of monitoring systems and applications to ensure efficient and reliable power delivery operations. \\ \hline
         
         Cyber-physical System & System in which entities (e.g. sensors, etc.) are connected to each other and to the internet through wired or wireless solutions. \\ \hline
         
         Man-in-The-Middle Attack (MiTM) & Cyber-attack during which a perpetrator eavesdrops the communication channel between two parties and often relays messages. \\ \hline
         
         Smart Grid & Modern power grid system with control and automation which enables bi-directional flow of electricity and communication in real-time. \\ \hline
         
         Supervisory control and data acquisition (SCADA) & Control system consisting of technological resources which enable high-level supervision of the power grid. \\ \hline
         
         Zero-day Vulnerability & Broad terminology describing a newly discovered form of vulnerability which is exploited by black hats to attack systems. \\ \hline
    \end{tabular}
    \caption{Terminology Description}
    \label{tab:my_label}
\end{table}

Following a brief introduction of the subject of interest of this manuscript in Section \ref{sec:Introduction}, we discuss, compare and contrast previous related survey articles against ours in Section \ref{sect:relatedsurvey}. The literature search methodology employed to gather, assess and select the relevant papers to our topic is presented in Section \ref{sec:litmethod}. Next, an overview of the cyber-physical security aspect of Active Distribution Systems along with some theoretical background of FDI attacks are highlighted in Section \ref{sec:background}. Section \ref{sec:proposedtaxonomy} provides a brief overview of the several categories of attack targets within our proposed taxonomy. Sections \ref{sec:enduserlevel}-\ref{sec:energybilling} discuss the suggested taxonomy of FDI threats, mainly from the adversarial point of view with reviews of previously undertaken studies. In particular, Section \ref{sec:enduserlevel} covers FDI attacks on the end user level, Section \ref{sec:fielddevices} explores similar attacks on field devices, Section \ref{sec:controlcentre} reviews those on the control center, while integrity threats on energy pricing and billing are covered under Section \ref{sec:energybilling}. Moreover, under Section \ref{sect:future researchgaps}, we identify some shortcomings of the current literature and provide some recommendations and directions for further research within this emerging field. Lastly, Section \ref{sect:conclusion} concludes this survey article.

\section{Related Works}

During the past couple of years, numerous amounts of work have surfaced on FDI threats in power systems. In this view, we first present related works in terms of FDI threats on power grids and how it can as well be extended to other fields. Secondly, we present a comparison of existing literature against ours to highlight the need for a new review.

\subsection{False Data Injection Threats}

The concept of FDI attacks was first coined by \cite{10.1145/1653662.1653666} and quickly became one of the stealthiest and devastating attacks on power systems. Due to the shift in paradigm from traditional power systems to smart grids, a complex system of interconnected sensors continuously gather data which are useful to ensure the safe and reliable operation of the grids. However, in recent years, the energy critical infrastructure has become an attractive attack target for malicious adversaries whereby they attempt to corrupt sensor readings to either disrupt the integrity or the availability of the system for their illicit gains \cite{7035067}. Traditionally, Bad Data Detectors (BDDs) have been employed to observe for corrupted information in power grid systems by comparing residual thresholds \cite{Sayghe_Hu_Zografopoulos_Liu_Dutta_Jin_Konstantinou_2020}. However, the speed of ongoing research throughout this field has brought about increasing stealthiness of such FDI attacks which tends to easily bypass BDDs. Following \cite{10.1145/1653662.1653666}'s research, throughout the last decade, a vast array of FDI attacks have been proposed in relation to disrupting several facets of the smart grid system. For instance, \cite{5622048} presented a stealth FDI attack on state estimation in a deregulated energy market setting which successfully bypassed BDDs and led to financial gains of the adversary during virtual energy bids. Another work by \cite{6503599} took a step further by proposing a more realistic FDI attack whereby the adversary conducted a gray-box attack with limited information in view of having similar impacts on the power systems. An interesting work proposed by \cite{6032057} enable adversaries to optimize their attacks by balancing the trade-off between increasing corruption of data and decreasing the probability of detection of the attack such that energy is leaked. Furthermore, \cite{6657769} investigated the effects of FDI attacks on real-time Locational Marginal Price (LMP). While, throughout the previous decade there has been several interesting FDI attacks proposed, within this manuscript, we will be mostly reflecting on FDI threats on Active Distribution Systems.

\subsubsection{FDI Threats in other fields}

Albeit the terminology of FDI attacks initially surfaced from smart grid use-cases \cite{7438916}, it is inherently pertinent to any other types of IoT-based systems as FDI attacks generally focus on the manipulation of data to impact the integrity of a system. While in the past, researchers have mostly concentrated on crafting FDI attacks and developing countermeasures in the power grid domain, some few recent literature \cite{Gu_Yu_Guo_Qiao_Guo_2021, 8247236} have extended FDI studies to other fields. For example, the work in \cite{8247236} proposed a novel attack construction and defense mechanism against FDI threats in Networked Radar Systems. Indeed, FDI attacks are relevant to critical healthcare infrastructure \cite{9162311}. Corrupted information injected in medical devices and sensors can pose imminent dangers to the health of individuals and can also cripple the healthcare industry of a particular region. For instance, medical devices consist of sensors that gather readings and health information about patients. The injection of corrupted sensor readings may lead to wrong medical diagnosis and thus jeopardize the well-being of a patient. Similarly, within the transportation infrastructure \cite{9623153}, the imputation of wrong information to sensor-based readings can eventually lead to several consequences such as traffic delays, accidents, etc. Furthermore, FDI attacks stand out as being one of the most aggressive threats on Industrial IoT (IIoT) systems where gross errors in sensor measurements can affect operations of machinery, etc \cite{9740499}. Moreover, there could be other instances and fields where FDI attacks could have potentially severe consequences such as in the finance sector whereby the manipulation of data could lead to misleading credit scores, in the military sector where drones can be hacked to inject false information which can trigger attacks to false targets, so on and so forth \cite{9162311}. Overall, it can be said that FDI threats, even though it has been mostly studied in the context of power systems, can be extended to several critical infrastructure and may cause adverse impacts in systems, all while remaining undetected for very long times. Therefore, it is highly necessary to have a better understanding of FDI attacks in the view of developing effective countermeasures. However, as aforementioned, within this manuscript, we shall be reflecting entirely on the threats targeting AMI-based distribution systems.  

\subsubsection{FDI Threats vs Other Types of Deception Threats}

Cybersecurity has now become a prominent aspect of Internet of Things due to the wide array of security attacks on IoT-based critical infrastructure. The wide array and heterogeneity of sensors present in IoT-based systems with each simultaneously sensing different types of data increases the complexity of the systems. Furthermore, due to the increased use of sensing devices in critical infrastructure, IoT-based devices are often produced at mass extent without taking into consideration security as a highest priority. Producers of Industrial IoT solutions in recent years have focused more on meeting demands rather that implementing strict digital doors to prevent unauthorized access which enables hackers with illicit intentions to carry out several types of deception attacks such as FDI attacks with devastating impacts. Deception attacks can be of several types e.g. Denial-of-Service (DoS) attacks, Trojan Attacks, Identity Deception, covert attacks and other miscellaneous deception threats \cite{Deception} . False Data Injection attack is one form of deception attack which targets data integrity issues. However, different types of deception threats have their own characteristics and impacts on a distribution system. For instance, DoS attacks target availability of systems while FDI threats are mostly related to data integrity issues. From an adversary's perspective, based on the CIA triad, an FDI attack tries to destabilize the integrity aspects of a distribution system while other deception attacks may have varying natures based on their own characteristics. To compare with a covert deception attack, FDI attacks attempt to disrupt the system within a short time interval for momentary gains while a covert attack allows an adversary to feed false data into a system such that the attack effects usually happen in the long-term \cite{7866869}. On the other hand, the main difference between other types of deception attacks is that deception threats may or may not be stealthy (e.g., DoS attacks are not stealthy) while FDI threats are categorized as stealthy attacks.

\subsection{Literature Review}
\label{sect:relatedsurvey}

\begin{table*}[]
    \centering
    \ding{51}: Included , \ding{55}: Not Included, \\ \ding{61}: Partially Included \vskip 0.06in
    \begin{tabular}{|p{1cm}|c|l|l|l|p{0.9cm}|p{0.9cm}|p{0.9cm}|p{0.9cm}|p{0.9cm}|p{0.9cm}|p{0.9cm}|p{0.9cm}|p{0.9cm}|}
    \noalign{\hrule height 0.5pt}
    \multicolumn{4}{|c} {\textbf{\textit{Comparison Attributes}}} & & \cite{6837157}& \cite{Mrabet_Kaabouch_Ghazi_Ghazi_2018} & \cite{7579185} & \cite{10.1504/IJWMC.2015.066756} & \cite{Liu_Li_2017} & \cite{7438916} & \cite{reda2021comprehensive} & \makecell{Our \\Paper}\\ \noalign{\hrule height 0.5pt}
 \multirow{19}{*}{\textit{\textbf{\makecell{Attack \\Target}}}} & \multicolumn{3}{c}{\textit{\textbf{End User Level}}} & & - & - & - & - & - & - & - & -  \\ \cline{2-13}
 & \multirow{3}{*}{\makecell{Energy\\Management}} & \multicolumn{2}{c}{Energy Storage} & & \ding{55} & \ding{55} & \ding{55} & \ding{51} & \ding{55} & \ding{55} & \ding{55} & \ding{51}\\ \cline{3-13}
 &  & \multicolumn{2}{c}{\makecell{Photo Voltaic \\Systems}} & & \ding{55} & \ding{55} & \ding{55} & \ding{55} & \ding{55} & \ding{55} & \ding{55} & \ding{51}\\\cline{3-13}
 &  & \multicolumn{2}{c}{\makecell{Smart Energy \\Management}} & & \ding{55} & \ding{55} & \ding{55}  & \ding{55} & \ding{55} & \ding{55} & \ding{51} & \ding{51}\\\cline{2-13}
 & \multirow{2}{*}{\makecell{Advanced \\Metering \\Infrastructure}} & \multicolumn{2}{c}{\makecell{Communication \\Networks}} & & \ding{55} & \ding{61} & \ding{55} & \ding{51}  & \ding{55} & \ding{55} & \ding{51} & \ding{51}\\\cline{3-13}
 &  & \multicolumn{2}{c}{\makecell{Smart \\Meters}} & & \ding{51} & \ding{61} & \ding{55} & \ding{51} & \ding{51} & \ding{55} & \ding{51} & \ding{51}\\\cline{2-13}
 & \multicolumn{3}{c}{\textit{\textbf{Field Devices}}} & & - & - & - & - & - & - & - & - \\ \cline{2-13}
 & \multicolumn{3}{c}{Voltage Regulators} & & \ding{55} & \ding{55} & \ding{55} & \ding{55} & \ding{55} & \ding{55} & \ding{55} & \ding{51}\\ \cline{2-13}
 & \multicolumn{3}{c}{Micro-PMU} & & \ding{55} & \ding{55} & \ding{55} & \ding{55} & \ding{55} & \ding{55} & \ding{55}& \ding{51}\\ \cline{2-13}
 & \multicolumn{3}{c}{Intelligent Field Devices} & & \ding{55} & \ding{55} &\ding{55} & \ding{55} & \ding{55} & \ding{55} & \ding{51} & \ding{51}\\ \cline{2-13}
 & \multicolumn{3}{c}{\textit{\textbf{Control Center}}} &  &  & & & & & & & \\ \cline{2-13}
 & \multicolumn{3}{c}{Volt-var Control} & & \ding{55} & \ding{55} & \ding{55} & \ding{55} & \ding{55} & \ding{55} & \ding{55} & \ding{51}\\ \cline{2-13}
   & \multirow{2}{*}{\makecell{Distribution\\State\\Estimation}} & \multicolumn{2}{c}{\makecell{Balanced\\ Single-phase}} & & \ding{55} & \ding{55} & \ding{51} & \ding{55} & \ding{55} & \ding{61} & \ding{61} & \ding{51}\\\cline{3-13}
 &  & \multicolumn{2}{c}{\makecell{Unbalanced\\ Multi-phase}} & & \ding{55} & \ding{55} & \ding{51} & \ding{55} & \ding{55} & \ding{61} & \ding{61} & \ding{51}\\\cline{2-13}
 & \multicolumn{3}{c}{\textit{\textbf{Energy Pricing \& Trading}}} & & - & - & - & - & - & - & - & - \\\cline{2-13}
 & \multicolumn{3}{c}{\makecell{Distribution Locational \\Marginal Pricing}} & & \ding{55} & \ding{55} & \ding{55} & \ding{55} & \ding{51} & \ding{55} & \ding{55}  & \ding{51}\\\cline{2-5}
 & \multicolumn{3}{c}{Real-time Pricing (RTP)} & & \ding{55} & \ding{55} & \ding{51} & \ding{55} & \ding{51} & \ding{55} & \ding{51} & \ding{51}\\ \cline{2-13}
 & \multicolumn{3}{c}{Transactive Energy Market} & & \ding{55} & \ding{55} & \ding{55} & \ding{55} & \ding{55} & \ding{55} & \ding{55} & \ding{51}\\ \cline{2-13}
 & \multicolumn{3}{c}{\makecell{Peer-to-peer Distributed \\Energy Trading}} & & \ding{55} & \ding{55} & \ding{55} & \ding{55} & \ding{55} & \ding{55} & \ding{55} & \ding{51}\\ \cline{1-13}
 \hline
\end{tabular}
\caption{Comparison of our survey against existing related surveys. \label{tab:comparison}}
\end{table*}

The authors in \cite{6837157} reviewed some of the early works on cyber threats in smart meters. The work by \cite{Mrabet_Kaabouch_Ghazi_Ghazi_2018} surveyed the attacks on smart grids and proposed a novel classification of cyber threats to smart grids based on methods used by hackers or penetration testers while compromising the grid. Furthermore, the authors in \cite{7579185} conducted a survey of data integrity attacks with respect to three major security aspects namely the construction of FDI attacks, the impacts of FDI attacks on state estimations for real-time electricity markets and lastly, defense mechanisms against those attacks. 

The work in \cite{10.1504/IJWMC.2015.066756} comprehensively surveyed FDI attacks with respect to power flow models namely Alternating Current and Direct Current. \cite{Liu_Li_2017} reviewed the existing literature based on several attack models, financial attack impacts and countermeasures within the transmission, distribution and micro-grid network. Similarly, research works in \cite{7438916} discuss the FDI attack models and their impacts on smart grid operations. Different to the previous work, \cite{reda2021comprehensive} surveyed and classified the related works on FDI attacks within all smart grids domains with respect to the attacks models, their attacks and the impacts of such attacks on grids.

As opposed to the existing related works, our manuscript attempts to present a thorough survey and review of the state-of-the-art data integrity attacks and develops a detailed taxonomy of aforementioned attacks with respect to points of attack across the modern distribution networks of power grids. A more detailed comparison of our paper against other surveys can be found in Table \ref{tab:comparison} above.

\section{Literature Review Methodology}

\label{sec:litmethod}
\begin{figure}
  \centering
  \includegraphics[width=8cm]{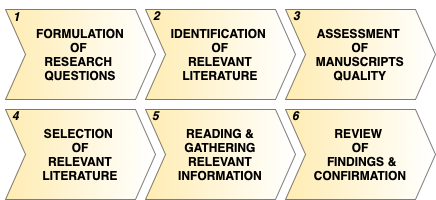}
  \caption{Survey Methodology employed (This figure describes a systematic methodology employed by our survey manuscript that incorporates six steps from formulating the research questions to review and confirmation of gathered findings). \label{gigsurmey}}
\end{figure}

This work covers a comprehensive survey and review of several independent studies on FDI threats in active distribution systems. Therefore, within this section, we provide an overview of our systematic literature search and selection process as shown in Figure \ref{gigsurmey}.

\subsection{Search Process}

Finding the relevant literature is vital to perform a comprehensive analysis of the topic. Therefore, we tackle this tedious search process using a structured methodology proposed by \cite{ref-misc2}. Relevant keywords and year filters are used to perform backward and forward searches on the academic databases to systematically identify high quality publications. Research databases used during our search process included IEEE Xplore, ACM Digital Library, Springer, Elsevier and others. The steps as shown in Figure \ref{gigsurmey} are then applied to each dataset using keywords such as "distribution system", "false data injection", "attacks", etc. 

\subsection{Literature Assessment \& Selection}

To ensure that our literature search does not consist of FDI articles related to other critical areas of research such as transportation, we restrict our search to smart grid distribution systems. Furthermore, all literature found from scholarly research sources deemed relevant to the subject in matter were manually evaluated against the scientific ranking platforms, namely SJR \footnote{SJR: Scientific Journal Ranking (Scimago)}$^{}$ \cite{sjrscimago} for journal articles and CORE\footnote{CORE: Computing Research and Education Association of Australasia}$^{}$ \cite{coreranking} for conferences, to consider prestigious and high quality studies. Following the confirmation of high quality literature, steps 5 \& 6 of Figure \ref{gigsurmey} were applied.

\section{Background}
\label{sec:background}
As part of the smart grid paradigm, ADSs are threatened by FDI attacks by potential adversaries. Therefore, in this section, we briefly give an overview of the cybersecurity aspect of active distribution systems followed by a theoretical background of FDI attacks.

\subsection{Cyber-physical Security of Active Distribution Systems}
As mentioned earlier on, the growing trend in renewable energy sources, predominantly installed on the distribution level, is gradually transforming the operations of distribution networks to an active paradigm \cite{Ghiani_Pilo_Celli_2018}. Contemporary innovative and intelligent technologies and devices are being used to revolutionize the once centralized, radial and "fit-and-forget" power distribution approach to a bi-directional automated scheme where efficiency and optimality of operations are guaranteed \cite{Pahwa_2015}. During the last few years, power system researchers around the globe have been significantly contributing to the shift from passive distribution systems to active ones. However, due to the conflicting nature of the objectives to be simultaneously optimized, research is still lagging behind to design economical, reliable and yet, cyber-resilient ADSs of the future \cite{Lakshmi_Ganguly_2018}. Therefore, throughout this section, we provide an overview of the main security goals and attacks on ADSs.

\subsubsection{Security Goals of Active Distribution Systems}
\begin{figure}
  \centering
  \includegraphics[width=9cm]{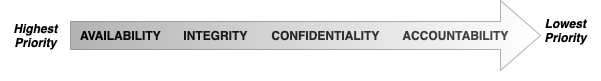}
  \caption{Order of Priority of four NIST Principles in relation to Active Distribution Systems. (The highest priority for ADSs is availability of the system while the lowest priority is accountability of the system).\label{NIST}}
\end{figure}

While utility operators are mainly focused on optimal and efficient distribution of energy to customers, security has taken the backseat and is now becoming a major concern. Therefore, it is important to extend and ensure the basic four NIST principles (refer to Figure \ref{NIST} of smart grid cybersecurity \cite{pricip2014} to ADSs, namely:
\begin{enumerate}[wide, labelwidth=!, labelindent=0pt]
    \item Availability: Availability is the most crucial security goal of ADSs to ensure uninterrupted supply of power to consumers at any given time. Countermeasures to protect ADSs against cyber-attacks must be within acceptable latency ranges while minimally impacting availability.
    \item Integrity: Being the second highest prioritized goal of ADSs cybersecurity, integrity must ensure that data is not illicitly altered and is from verifiable sources. The alteration and/or destruction of the true data leads to a loss of data integrity. Indeed, as data integrity degrades, it negatively impacts the reliability ad stability of the distribution system.   
    
    \item Confidentiality: While confidentiality may seem to be of lesser importance in ensuring reliability of distribution networks, vulnerable Advanced Metering Infrastructure (AMI) must be protected against unauthorized leakage of private customer or proprietary information.
    \item Accountability: The last security objective being accountability, relates to consumers being responsible for their actions such as during billing, consumption, etc. 
\end{enumerate}

\subsubsection{Cyber-physical Attacks on Active Distribution Systems}

While Distributed Generation (DG) is a viable solution to sustain the exponential growth in load demand, the integration of converter-based DG units including Photovoltaic (PV) generators deteriorates the power quality through the injection of harmonics to the distribution network \cite{Lakshmi_Ganguly_2018}. Coupled with the integration of non-standardized DG units, the addition of contemporary technologies for efficient and automated power distribution management greatly increases the complexity of distribution networks \cite{Jokar_Arianpoo_Leung_2016}. This opens up several security vulnerabilities which can be exploited by potential adversaries for both financial and political gains. 

\subsection{False Data Injection Threats}

Dubbed as one of the most critical malicious cyber-threats in power systems, an adversary tends to deliberately compromise sensor readings by orchestrating injection of false information into sensor measurements. Specifically, during FDI attacks, an attacker attempts to compromise sensor readings stealthily such that gross errors are introduced into data or aggregation procedures while evading detection. The objective of an attacker is to introduce an attack vector $a$ into the data measurements while evading bad data detection by operators. This results in maliciously compromising the state variables across distribution networks. In simple words, during a FDI attack, an adversary manipulates the real measuring vector which can be mathematically denoted as $z_{a} = z + a$ where $z$ is the original sensor information, $z_{a}$ is the corrupted measurement and $a$ is the attack vector.  In any case, corrupted information can be achieved by one of the following: 1) Deletion of data from the original measurement information, $z$. 2) Change of data in the original measurement information, $z$. 3) Addition of fake data to the original measurement information, $z$.  Furthermore, adversaries may be subject to several inequality constraints that simultaneously minimize the probability of detection of the launched attack and minimize the probability of detection of such attacks.

\begin{figure}
  \centering
  \includegraphics[width=8cm]{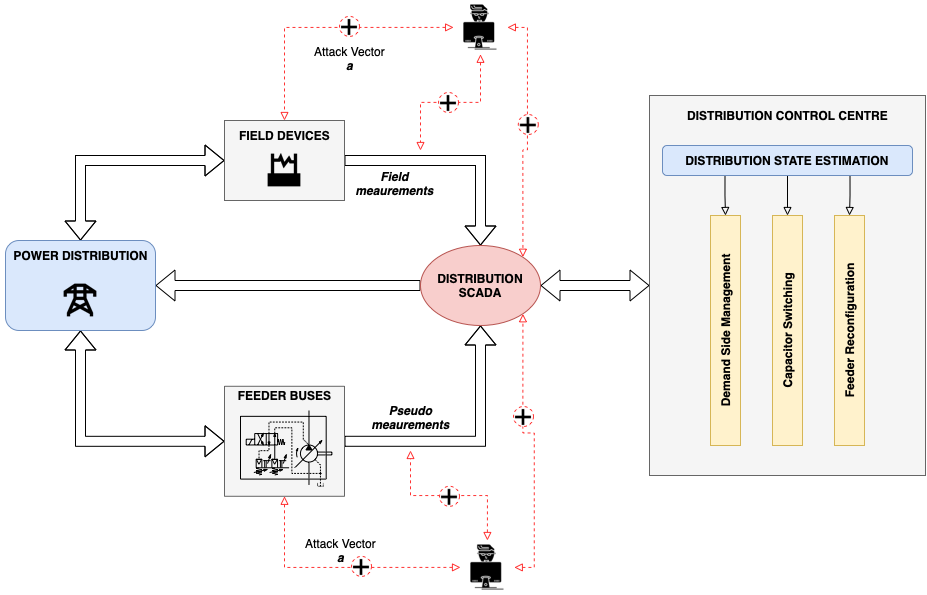}
  \caption{Overall architecture of FDI attacks in a connected power distribution system grid. (While the information flow throughout a distribution system is depicted, an adversary can inject corrupted information at the field and pseudo measurements, in the communication channels as well as in the SCADA systems as main points of attacks.) \label{intrinsic}}
\end{figure}

Based on the security goals discussed earlier, integrity is the second most prioritized security goal whereby a distribution system must ensure the accuracy of its information. However, an FDI attack is a class of data integrity threat that exploits sensor (e.g., smart meters of a distribution system) measurements through the injection of corrupted information. Traditionally, BDDs have been employed to detect corrupted sensor readings by calculating the residual between the measured and estimated system states which are obtained from the state estimation process. Due to intelligent mechanisms used by adversaries to choose the attack vector, the residual measurement becomes very close to the regular normal behaviour without any FDI threat. Hence, such type of attack remains stealthy which adversely affects the integrity of the distribution system. As sensor data plays a critical role in different key operational modules \cite{Sun_Hahn_Liu_2018}, such corruption eventually impacts the operational decisions of distribution systems. As shown in Figure \ref{intrinsic}, we provide an overall architecture of FDI threats in a connected smart grid which highlights the intrinsic relationship of FDI threats on the integrity and reliability of the distribution system.

As aforementioned, with latest developments within this field, FDI threats have become more stealthy such that now they are able to completely circumvent BDDs as well as other proposed countermeasures. Therefore, to preserve the stealthiness of FDI threats, the attack vector $z_{a}$ must remain in the boundary conditions, $z_{min}$ and $z_{max}$. For any FDI attack vector, $z_{a}$, it should circumvent BDDs if 
\begin{equation}
    z_{a} = z + a, z_{min} \leq z_{a} \geq z_{max}
\end{equation}

where $z_{min}$ and $z_{max}$ is assumed to be known by the adversary \cite{https://doi.org/10.48550/arxiv.1910.01716}. 

\section{Proposed Taxonomy}
\label{sec:proposedtaxonomy}
\begin{figure*}
  \centering
  \includegraphics[width=\linewidth]{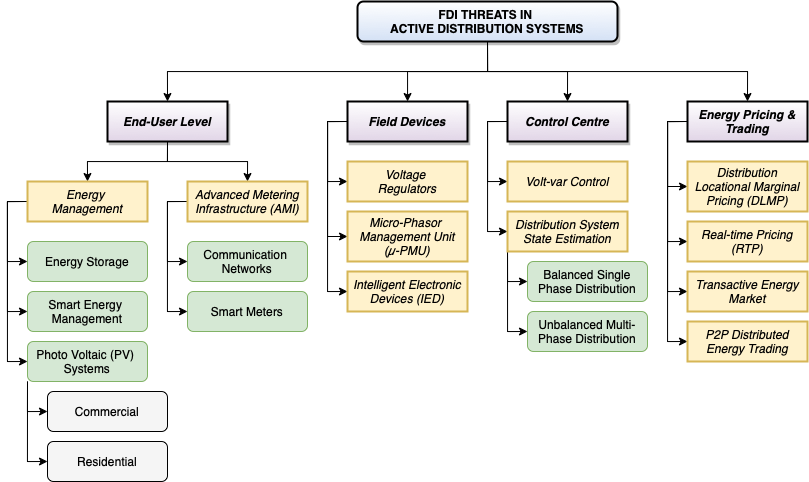}
  \caption{Taxonomy of threats on AMI-based distribution systems. \label{figuretaxonomy}}
\end{figure*}

Several studies on FDI threats to distribution systems have been conducted by researchers. In this survey, we propose a taxonomy to classify threats on ADSs with respect to attack targets as shown in Figure \ref{figuretaxonomy} below. The four major attack target levels are:

\begin{enumerate}[wide, labelwidth=!, labelindent=0pt]
    \item End-user Level: Edge devices based at the end-user side/customer side are often prone to several cyber-threats due to the lack of standardization and secure communication protocols. Therefore, we address the related FDI threats within this attack target level in Section \ref{sec:enduserlevel}.
    
    \item Field Devices: The integration of IoT-based devices increases the complexity of the wireless field area networks which opens up security vulnerabilities and can impact the grid. We discuss the FDI threats relevant to this category in Section \ref{sec:fielddevices} 
    
    \item Control Center: Network operation outsourcing and the complexity of connection and communication of Supervisory Control and Data Acquisition (SCADA) components within the power distribution control center are prone to FDI threats which can adversely impact the stable operation of smart grids. FDI threats targeting the control center are discussed in Section \ref{sec:controlcentre}.
    
    \item Energy Pricing \& Trading: The shift to a modern decentralized supply and demand side management approach due to power system restructuring and the integration of renewable power sources threatens the energy billing and trading process for the adversaries' malicious financial gains. We therefore address this category in Section \ref{sec:energybilling}.
    
\end{enumerate}

\section{Threats on End-User Level}
\label{sec:enduserlevel}

The increase in complexity along with the bi-directional communication provoked by renewable energy management systems and advanced metering infrastructure at the end-user level opens up several security vulnerabilities within ADSs \cite{Rashed_Mohassel_Fung_Mohammadi_Raahemifar_2014}. In this view, we present the FDI threats on end-user level as in Tables \ref{tab:enduserthreats} and .\ref{tab:enduseramithreats}.

\subsection{Energy Management}
The traditional power grid has shifted from a current centralized power generation paradigm to a distributed one resulting from the integration of local sustainable power generation systems \cite{6450343}. To balance the increasing demand for power, operators are deploying renewable energy sources from different vendors which increases the complexity of power grids \cite{Balezentis_Streimikiene_Mikalauskas_Shen_2021}. Such increased complexity increases the distribution systems to severe attacks that may disrupt the operations of the grid. In this view, we present the state-of-the-art threats on energy management within distribution systems on three sub-categories namely:

\subsubsection{Energy Storage}
Renewable energy generated from decentralized production do not often provide immediate response to demand and therefore requires energy storage usually in the form of batteries. The integration with several state-of-the-art technologies including IoT and cloud computing is increasing the complexity of energy management systems in the distribution side. This increased complexity is however exposing energy management systems to severe cyber threats. Zhuang and Liang \cite{9284586} proposed the static analytical injection of corrupted State-of-Charge (SoC) information with small magnitude into weighted least squares (WLS)-based state estimators. Experimental validations revealed that their formulated static sequential data integrity attack drastically affected the accuracy of the SoC estimation by 17\% while still being able to circumvent state-of-the-art  measurement residual-based bad data detection algorithms and innovation test.

\begin{table*}[!h]
\begin{tabular}{ll|p{2.5cm}|p{2.5cm}|p{1.5cm}|p{3.5cm}|}
\hline
\multicolumn{2}{|l|}{\textbf{Category}} & \textbf{Ref No} & \textbf{Attack Target}           & \textbf{Attack Type} & \textbf{Attack Mechanism}\\ \hline
\multicolumn{2}{|l|}{Energy Storage} & \cite{9284586} & Battery Terminal Voltage & Sequential data injection & Constraint optimization based on Coulomb Counting Method to inject attack vectors post-attacking. \\ \hline

\multicolumn{1}{|l|}{\multirow{5}{*}{\makecell{Photovoltaic\\ Inverters}}} & \multirow{3}{*}{Commercial}  & \cite{OlowuFDIInvest2020}                                                                  & Volt-VAR, Volt-Watt \& Constant power factor of Smart Inverters & Short-term data injection.                         & Gross data is injected to change the set-points based on smart inverter settings before and after the attack.                      \\ \cline{3-6} 
\multicolumn{1}{|l|}{}                                        &                              & \cite{Tertytchny_Karbouj_Hadjidemetriou_Charalambous_Michael_Sazos_Maniatakos_2020} & PV penetration levels data packets                              & MiTM and Short-term data injection                 & Analysis of collected packets in order to inject corrupted measurements which will overfeed the Smart Inverter and cause tripping. \\ \cline{3-6} 
\multicolumn{1}{|l|}{}                                        &                              & \cite{BaruaFaruqueUSENIX2020}                                                              & Hall sensor measurement                                         & DoS                                                & Non-invasive physical magnetic spoofing technique with adversarial control.                                                        \\ \cline{2-6} 
\multicolumn{1}{|l|}{}                                        & \multirow{2}{*}{Residential} & \cite{Kandasamy_2020}                                                                     & Reactive Power information                                      & Coordinated MiTM attack                            & Physical Tampering with the actuation command of inverters.                                                                        \\ \cline{3-6} 
\multicolumn{1}{|l|}{}                                        &                              & \cite{lindstrom2021power}                                                                  & Power penetration data                                          & Power Injection Attack                             & Constraint Optimization of cause maximal voltage deviation with an attack at one node in the finite time interval                  \\ \hline
\multicolumn{2}{|l|}{\multirow{3}{*}{\makecell{Smart Energy \\Management System}}}                        & \cite{8586588}                                                                             & Load \& Pricing data                                            & DoS, Phishing attacks \& short-term data injection & Genetic Algorithm Optimization to inject corrupted information into load profiles.                                                 \\ \cline{3-6} 
\multicolumn{2}{|l|}{}                                                                       & \cite{9256449}                                                                             & Price data                                                      & Long-term data injection                           & Genetic Algorithm Optimization for minimizing the total electricity cost drawn from grid                                           \\ \cline{3-6} 
\multicolumn{2}{|l|}{}                                                                       & \cite{Sethi_Mukherjee_Singh_Misra_Mohanty_2020}                                       & Price data  & MiTM long-term data injection                      & Bi-level linear programming Optimization \\ \hline
\end{tabular}
\caption{Comparative View of threats on End-User Level - Energy Management \label{tab:enduserthreats}}
\end{table*}

\subsubsection{Photovoltaic Inverters}

The rapid technological integration of smart photovoltaic inverters with Distributed Energy Resources (DERs) coupled with environmental sustainability objectives has led to the proliferation of inverter-based Distributed Energy Resources (IBDERs) in electric power grids \cite{Yazdaninejadi_Hamidi_Golshannavaz_Aminifar_Teimourzadeh_2019}. However, the successful deployment of photovoltaic inverters is still prone to security and privacy breaches which may have devastating implications on distribution power systems \cite{Wankhede_Paliwal_Kirar_2020}. In this view, we present the related recent state-of-the-art literature on threats against smart solar inverters within two application scenarios namely: 

\begin{enumerate} [wide, labelwidth=!, labelindent=0pt]
    \item Commercial Domain: \cite{OlowuFDIInvest2020} first investigated the impact of data integrity attacks to a commercial distribution feeder by injecting false data to the three most common Smart Inverters functionalities namely Volt-VAR, Volt-Watt and Constant Power factor (CPF). Experimental evaluation of their proposed attack revealed that the attacks severely impacted the voltage profile and the reactive power injection from the capacitor. Similarly, \cite{Tertytchny_Karbouj_Hadjidemetriou_Charalambous_Michael_Sazos_Maniatakos_2020} proposed a man in the middle attack to overload a targeted feeder by injecting false data to all packets between the smart meter and the ancillary services controller. This trips the overcurrent protection relay and in turn leads to a regional blackout. After conducting further risk analysis, it was revealed that the efficacy of their proposed attack rose drastically with increasing solar photovoltaic inverter capacity. On the other hand, \cite{BaruaFaruqueUSENIX2020} crafted a non-invasive DoS attack whereby an adversary injects  false measurement data into hall sensors of solar inverters through magnetic spoofing. The data integrity attack further propagates to compromise the whole inverter eventually which may cause grid instability and failures. As opposed to the works presented in \cite{OlowuFDIInvest2020} and \cite{Tertytchny_Karbouj_Hadjidemetriou_Charalambous_Michael_Sazos_Maniatakos_2020}, the false data injection in this case comes from physical domains by exploiting the external magnetic fields.
    
    \item Residential Domain: \cite{lindstrom2021power} investigated the consequences of a deceptive power injection attack against the physical layer of a smart distribution grid with radial topology. The adversary tends to maximize the voltage deviation which impacts the inverter and eventually shuts down part of a grid. \cite{Kandasamy_2020} demonstrated the effect of bias attacks in prosumer-based reactive power control. The goal of the attacker is to inject gross errors to the actuation commands of smart photovoltaic inverter which drastically reduces its voltage and increases the current flow. The over-current leads to thermal tripping of the inverter and  as a result of which, a regional blackout occurs. 
\end{enumerate}

\subsubsection{Smart Energy Management System}

The wide use of smart residential components and integration of IT has revolutionized residential homes into smart ones. Further coupled with the incorporation of the two-way communication with smart grids and advanced intelligence in exchange for economic benefits, Smart Home Energy Management Systems (SHEMSs) can be considered as systems that provide optimal energy management services in view of efficiently monitoring and managing electricity generation, storage, and consumption in smart residential homes \cite{SonMoon2010, JinsooChoi2011}. However, SHEMSs are becoming more prone to cyber threats which may have devastating impacts. \cite{8586588} first investigated the effects of cyber attacks on SHEMSs by launching DoS and phishing attacks in view of modifying the load profile data and the dynamic pricing information. The researchers highlighted that such types of attacks can temporarily disrupt the scheduling operation and can adversely affect the energy pricing. Similarly, \cite{9256449} proposed a pricing attack on smart homes under a third party aggregator system at different attack points. The authors concluded that the vulnerability of SHEMSs to pricing attacks will impede its adoption in smart homes. In similar line, \cite{Sethi_Mukherjee_Singh_Misra_Mohanty_2020} proposed the injection of corrupted pricing data to disrupt scheduling and pricing operations. During this particular attack, the attacker, in perspective of a customer, plans to decrease the price of the electricity bill from being originally at $\$2.11$ to $\$1.79$ and the grid power import from $78.15$kW to $76.99$kW. While this difference in pricing seems slight, corrupting electricity bills may be profitable to a customer in the long run and cause financial losses for the electricity utilities. For instance, depending on how many households have increased their consumption, the collective  energy consumption may be considerable particularly if it happens in an unwanted period of time such as peak consumption time.   

\subsection{Advanced Metering Infrastructure (AMI)}

\begin{figure}
  \centering
  \includegraphics[width=8cm]{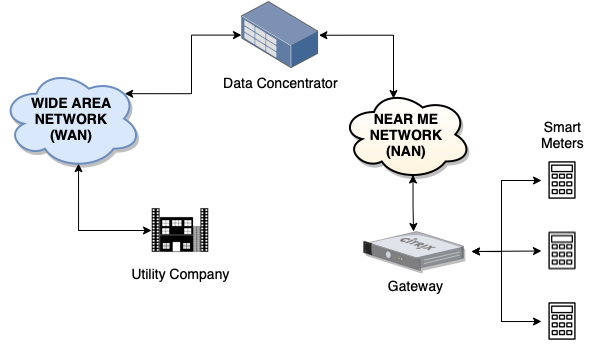}
  \caption{Block diagram providing an overview of the components and networks within an Advanced Metering Infrastructure (AMI). The figure also depicts the bi-directional communication between the utility providers and the consumers.\label{AMI}}
\end{figure}

Advanced Metering Infrastructure (AMI) is now regarded as the backbone of smart grids enabling real-time communications between utility providers and customers \cite{Mohassel_Fung_Mohammadi_Raahemifar_2014}. However, the increasing complexity of AMIs has significantly resulted in a rise in the number of cyber-threats on AMIs in recent years. The highly sensitive data sensed and transmitted within the AMI (as depicted in Figure \ref{AMI} has empowered adversaries to exploit the vulnerable points of an AMI.  In this view, we present the related recent state-of-the-art literature on threats against AMIs at the two vulnerable points of entry namely: 

\begin{table*}[]
\begin{tabular}{|l|p{2.5cm}|p{2.5cm}|p{2cm}|p{3.5cm}|}
\hline
\textbf{Category}                       & \textbf{Ref No}                                                 & \textbf{Attack Target}                                   & \textbf{Attack Type}           & \textbf{Attack Mechanism}                                                                                                                   \\ \hline
\multirow{5}{*}{Smart Meters}           & \cite{Lo_Ansari_2013}                        & Energy Profile                                           & MiTM short-term data injection & Dynamic programming optimization of the attack formulation to a coin change problem.                                                        \\ \cline{2-5} 
                                        & \cite{Khanna_Panigrahi_Joshi_2016}          & Smart meter energy generation data                       & Short-term data injection      & Constraint Optimization to maximize the power injection of a bus.                                                                           \\ \cline{2-5} 
                                        & \cite{Fan_Li_Cao_2017}                      & Load Profile                                             & Privacy attack                 & Application signature extraction \& identification.                                                                                         \\ \cline{2-5} 
                                        & \cite{Wu_Chen_Weng_Wei_Li_Qiu_Liu_2019} & Energy Consumption Data                                  & False load attack              & Sending periodic circuit-OFF signals to the IGBT gate such that the circuit is switched off when the meter is sampling the current reading. \\ \cline{2-5} 
                                        & \cite{Ismail_Shaaban_Naidu_Serpedin_2020}  & Smart meter energy generation data                       & Short-term data injection      & The adversary manipulates their readings to claim higher supplied energy to the grid and hence falsely overcharge the utility company.      \\ \hline
\multirow{2}{*}{\makecell{Communication \\Networks}} & \cite{Yi_Zhu_Zhang_Wu_Li_2014}            & Communication packets between smart meters and utilities & Puppet DDoS attacks            & An adversary floods a puppet node with data packets so as to exhaust the network communication bandwidth and node energy.                   \\ \cline{2-5} 
                                        & \cite{Boudko_Abie_2018}                      & Communication messages                                   & MiTM Short term data injection & Evolutionary game theory.                                                                                                                   \\ \hline
\end{tabular}
\caption{Comparative View of threats on End-User Level - Advanced Metering Infrastructure \label{tab:enduseramithreats}}
\end{table*}

\subsubsection{Smart Meters}

Energy theft has always been a major concern faced by utility companies throughout the globe and dates as far as the late 1800s \cite{Combatingenergytheftwithanalytics}. Physical interventions through illegal connections and meter tampering contribute to significant revenue losses of utility providers \cite{Czechowski_Kosek_2016}. While the introduction of smart meters has brought along an array of opportunities including accurate load forecasting, network controllability, etc., researchers have been investigating new attacks to compromise smart meters. \cite{Lo_Ansari_2013} proposed a combination sum of energy profiles (CONSUMER) attack whereby an adversary reduces its own energy consumption by injecting false data into its own smart meter. Furthermore, to lower the changes of fraud detection by the utility company, the attacker compensates the discrepancy in measurement by injecting corrupted data into least possible number of other smart meters within the neighborhood area network. The authors inferred that several machine learning detection schemes will indeed fail to detect such alterations, especially if the adversary injects corrupted data of small magnitude. \cite{Khanna_Panigrahi_Joshi_2016} developed a new attack for modifying the system state to portray false increased energy exports by injecting false data into smart meters at generator buses. The authors claim that their attacks were successful at enabling an adversary to gain momentary economic gains whilst attacking the least number of smart meters. The researchers in \cite{Fan_Li_Cao_2017} focus on the exploitation of reactive power data from smart meters to infer power consumption of home appliances by initially extracting a one-minute window of the reactive power waveform to capture the essential characteristics of appliances, filtering deceptive events, detecting real events and lastly identifying the appliances. Evaluation results on  real residential power consumption data revealed that such attacks are highly effective for violating the privacy of residents. \cite{Wu_Chen_Weng_Wei_Li_Qiu_Liu_2019} proposed the closing and opening of a power main line (or its branch) synchronous to the rate of sampling of the smart meter by injecting corrupted signals to an insulated gate bipolar transistor. After validation, this attack is found to be immune to all standard security countermeasures as well as very effective in significantly reducing power consumption bills. \cite{Ismail_Shaaban_Naidu_Serpedin_2020} introduced a setting whereby the malicious customers hack their smart meters and increase the solar power generation readings using several types of cyber attacks namely partial increment attacks, minimum generation attacks and peak generation attack which resulted in the overcharging of utility companies. 

\subsubsection{Communication Networks}

Real-time two-way communication is of crucial importance in AMIs. As aforementioned, a high volume of extremely sensitive data is transmitted to and from the utility provider and the end-user \cite{Mohassel_Fung_Mohammadi_Raahemifar_2014}. Nonetheless, the numerous advantages brought about by the two-way communication also increases the vulnerability of an AMI network to malicious attacks.  \cite{Yi_Zhu_Zhang_Wu_Li_2014} introduced a novel type of Denial-of-Service attack known as puppet attacks on AMI networks by flooding a puppet node with adversarial route request packets so as to exhaust the bandwidth of communication and the node energy. Experimental evaluations show that their proposed attack significantly decreases the performance of the AMI communication network and the packet delivery rate reduces from 20\% to 10\%. The authors in \cite{Boudko_Abie_2018} proposed a one-shot evolutionary game theoretical framework to model data integrity attacks on AMIs nodes to allow adaptive selection of strategies under resource constraints such that the node pay-offs are maximized. Results highlight that adversaries prefer nodes with higher aggregation and the attacker uses most of his budget to attack the Head-End System node.

\section{Threats on Field Devices}
\label{sec:fielddevices}

With the increased deployment of field devices along distribution feeders and within substations, grids are now able to smartly and efficiently perform distribution automation, automatic load shedding, outage management, etc \cite{Chhaya_Sharma_Kumar_Bhagwatikar_2018}. However, the implementation of several such IoT-based devices increases the complexity of the wireless field area networks and therefore, exposes several security vulnerabilities which can be exploited by adversaries. In this view, we survey the state-of-the-art threats on field devices as in Table \ref{tab:threatsfield}:

\begin{table*}[]
\begin{tabular}{|p{2cm}|p{3cm}|p{3cm}|p{2.5cm}|p{3.5cm}|}
\hline
\textbf{Main Category} & \textbf{Ref No.}  & \textbf{Attack Target} & \textbf{Attack Type} & \textbf{Attack Mechanism} \\ \hline
\multirow{3}{*}{\textbf{\textit{\makecell{Voltage \\Regulators}}}} & \cite{7007753} & Load ratio control switch sensor measurement data & Short term data injection & Constraint Optimization which enables the attacker to suppress/ induce tap changes. \\ \cline{2-5} 
 & \cite{7301476}  & Bus Voltage Measurement & Voltage reference attack \& Voltage measurement routing attack & During the first attack, the adversary inputs corrupted information to the bus voltage measurements while during the routing attack, the perpetrator redirects voltage measurements to another receiving bus in the network. \\ \cline{2-5} 
 & \cite{Ma_Teixeira_van7}  & Bus Voltage Measurement & Short-term data injection & Multiplicative bounded scaling factor for crafting attacks to one node only. \\ \hline
\multirow{2}{*}{\textbf{\textit{\makecell{Intelligent \\ Electronic \\ Devices (IEDs)}}}} & \cite{6396770}  & Electromagnetic Field & Electromagnetic Threats & Electromagnetic Disturbances that may either be produced deliberately or is naturally occurring. \\ \cline{2-5} 
 & \cite{8097030}  & Faulty IEDs & Implementation attacks (malicious fault injection attacks \& hardware Trojan) & \_ \\ \hline
\multirow{2}{*}{\textbf{\textit{\makecell{Micro-Phasor \\ Measurement \\ Units ($\mu$-PMU)}}}} & \cite{8566281}  & Voltage \& Angle measurements & Short-term data injection & High Value FDI attack \\ \cline{2-5} 
 & \cite{9219167}  & Phase angle channel measurements & Unsynchronized \& event-synchronizedattacks & During the first attack, the attacker is unable to synchronize the FDI attack with the occurrence of pre-event and post-event measurements as opposed to during event-synchronized attacks. \\ \hline
\end{tabular}
\caption{Comparative View of threats on Field Devices.}
\label{tab:threatsfield}
\end{table*}
\subsection{Voltage Regulators}

The multi-directional power flow achieved from the integration of DERs along with the additional stress on voltage control devices caused by the stochastic and concentrated power profiles of Plug-in Electric Vehicles (PEVs) can lead to over voltages, under voltages, high system losses, excessive tap operations and so on \cite{8231977}. Therefore, grid operators employ voltage regulation devices such as on load tap changers (OLTC), ratio control transformers (LRTs), Step Voltage Regulators (SVRs) and shunt capacitors to mitigate the previously mentioned issues. However, the centralized nature of voltage regulation enables attackers to create bottom-to-top attacks propagation which can ultimately lead to severe outages \cite{Sun_Hahn_Liu_2018}. Therefore, \cite{7007753} proposed the falsification of a limited number of sensor measurements through suppressing or inducing tap changes at the LRT to maximize overvoltage or undervoltage violations by an adversary with full knowledge of the control algorithm. Simulation results on a distribution network with one feeder modeled at a smaller scale from a residential district in Japan with real-world data revealed comparable results with two upward tap changes each for cases with and without PVs which result in undervoltages between 1.00 to 7.86 $\times 10^5$at some nodes. The work in \cite{7301476} considered two types of attacks on known as Voltage Reference Attack (VRA) where an adversary injects false-data into the communication network and Voltage Measurement Routing Attack (VMRA) where an adversary redirects data to a wrong receiver bus within a network by manipulating reference signals. The impact of the attacks were further characterized based on control-theoretic tools namely stability and input-output induced norm of linearized systems. Numerical simulations resulted in a step change in the voltage profile ranging between 0.5\% to 8\% at the buses during VRA and between $-3$\% to 8\% during VMRA.  \cite{Ma_Teixeira_van7} extended the work in \cite{7301476} by mostly focusing on the manipulation of sensor measurements with similar approaches to characterize the impact of the attacks. Experimental evaluations of an islanded four-bus power distribution network with a line topology reveals that the closed-loop system  under attack is asymptotically stable and the voltage deviation relates to falsification ratio, $\delta$, with an exponential decrease of 90\% from  $\delta = 0$ to $\delta = 0.5$ and a near linear increase of 30\% from $\delta = 0.5$ to $\delta = 1$. Such attack whereby the adversary decreases the voltage measurement received by the droop controller has higher impacts on the neighboring nodes within a line network. 

\subsection{Intelligent Electronic Devices}

Intelligent Electronic Devices (IEDs) are widely used to enhance automation within smart substations \cite{8006250}. However, the vast spatial complexity and the complex management hierarchy opens up potential vulnerabilities that can be easily exploited by adversaries \cite{8542059}. Radasky and Hoad \cite{6396770} studied the impacts of three High Power Electromagnetic (HPEM) threats on IEDs namely Intentional Electromagnetic Interference (IEMI) whereby an adversary deliberately increases the electromagnetic disturbances, High Altitude Electromagnetic Pulse (HEMP) whereby an attacker create a 30 km high-altitude nuclear burst that produces intense elctromagnetic signals which reach the earth and lastly, Extreme Geomagnetic Storms which is a natural disaster that cause a significant rapid distortion of the geomagnetic field at the earth's surface. A fast rising and short $2.5/2.5$ ns electric field pulse during Early-time HEMP results in levels of the order of 20kV to IEDs and can even trip protective relays at substations. Similarly, other types of attacks distort and damage IEDs present in sub-stations. \cite{8097030} studied the effects of two types of implementation attacks namely malicious low-cost fault injection attack by underfeeding the micro-controller and hardware Trojan attack on protective distribution relays in smart sub-stations. Simulations on ARMv7-based micro-controller with 100 attack executions revealed a significant increase in the reaction time to a trip signal of up to 9 times. Such attacks may eventually cause delays in tripping part of a power distribution network. Any delays above a threshold may cause drastic damage to some equipment including power network assets such as power lines, transformers, etc.

\subsection{Micro Phasor Measurement Units}

The shift from a passive to an active distribution system has overseen rapid developments in Micro Phasor Measurement Units ($\mu$-PMU) to guarantee real-time and accurate synchronized phasor data measurements of electricity including voltage, current, and frequency \cite{7955020}. However, the ubiquitous nature of $\mu$-PMU increases the attack surface of ADSs to adversaries \cite{Shukla_Dutta_Sadhu_2021}. Ren and Jordan \cite{8566281} proposed the injection of corrupted high value data in $\mu$-PMU measurements (voltage and angles) to assess the robustness of Weighted Least Absolute Value and Weighted Least Squares (WLS) estimators. Simulations on an IEEE 37 Node Test Feeder with $\mu$-PMU buses revealed a stunning 96.7\% error in the measurement which in turn results in the divergence of the WLS estimator. Furthermore, the study undertaken in \cite{9219167} developed two types of false data injection threats namely event-unsynchronized attacks and event-synchronized attacks on $\mu$-PMUs. During the event-unsynchronized attack, the attacker distorts the data at the magnitude channel or the phase angle channel of the micro-PMU while being unable to synchronize the attack with the pre-event phasor measurements and the post-event phasor measurements. On the other hand, during an event-synchronized attack, the adversary may compromise the pre-event phasor measurements and the post-event phasor measurements, separately which can easily bypass bad data detectors and be triggered only during event occurrences. Experiments on an  IEEE-33 bus test system followed by geometric analysis revealed that event-synchronized attack require 20 times lesser injection of error measurements than their event-unsynchronized counterparts and causes higher impacts with a 50 times smaller fractional change in phasor angle and voltage magnitude while still remaining stealthy. Such a targeted attack could be limited in scope but result in a major impact on the operation of the power grid by highly deviating the outcome of the event-based methods.

\section{Threats on Control Center}
\label{sec:controlcentre}

\begin{figure}
  \centering
  \includegraphics[width=8cm]{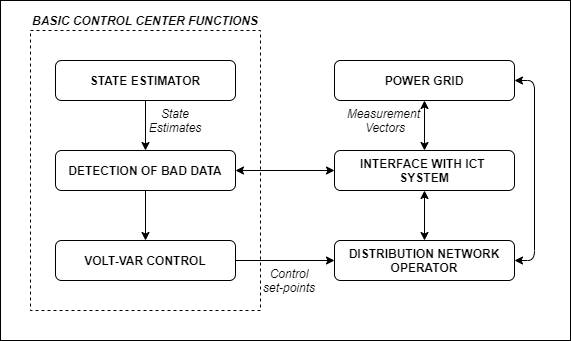}
  \caption{Block diagram depicting a typical distribution control center system. (The figure represents the basic functionalities performed by a control centre as well as the flow of information within a distribution system.)}
\end{figure}

Following the Northeast blackout of 2003 \cite{Blackout2003}, traditional power grids have been revamped with the integration of latest technologies which enables grid operators to have more control and monitoring over the distribution system \cite{Operation_Centers_2019}. However, network operation outsourcing and the complexity of connection and communication of Supervisory Control and Data Acquisition (SCADA) components within power distribution control systems opens up security vulnerabilities which can be exploited for financial or political welfare \cite{Tom_Sankaranarayanan_2017}. Therefore, we present a taxonomy of threats on the power distribution systems control center as in Table \ref{tab:threatscontrol}:

\begin{table*}[]
\begin{tabular}{|p{1.5cm}|p{1.5cm}|p{2.5cm}|p{2.5cm}|p{1.5cm}|p{4.5cm}|}
\hline
\textbf{Main Category} & \textbf{Sub Category} & \textbf{Ref No.} & \textbf{Attack Target} & \textbf{Attack Type} & \textbf{Attack Mechanism} \\ \hline
\multirow{4}{*}{\textbf{\textit{\makecell{Volt-var\\ Control}}}} & \multirow{4}{*}  & \cite{6859265}  & Voltage node measure- ments & MiTM Stealth attack & Addition of Arbitrary voltage while subtracting attack vector from capacitor configurations. \\ \cline{3-6} 
 &  & \cite{JuAdversarialDistributedVoltage}   & Reactive power of DER devices & Short Term Data Injection & Topology-agnostic approach with constraint optimization \\ \cline{3-6} 
 &  & \cite{8667289} &   Distribution feeder voltage profile & Short Term Data Injection & Bilevel optimization problem using mixed integer linear programming \\ \cline{3-6} 
 &  & \cite{Shen_Liu_Xu_Lu_2021} &  Volatge \& load information & Load redistribution attack & Single-leader-multi-follower bi-level mixed-integer linear Optimization \\ \hline
\multirow{3}{*}{\textbf{\textit{\makecell{State \\Estimation}}}} & Balanced Single-Phase Distribution & \cite{8307441} &   Meter measurement information & Short Term data injection & Non-linear attack policy based on weighted least squares optimization \\ \cline{2-6} 
 & \multirow{2}{*}{\makecell{Unbalanced \\Multi-phase \\Distribution}} & \cite{8625609} &   Bus voltage measurements & Short Term data injection & Constraint optimization based on proposed local state-based linear DSSE \\ \cline{3-6} 
 &  & \cite{9035401} &   Smart Meter Data & Load redistribution attack & Bi-level optimization problem transformation to a single-level optimization problem based on Karush-Kuhn-Tucker conditions of the lower level optimization problem \\ \hline
\end{tabular}
\caption{Comparative View of threats on Control Center.}
\label{tab:threatscontrol}
\end{table*}
\subsection{Volt-var Control}

Distribution Automation Systems (DASs) have emerged as effective solutions to improve operational efficiency of distribution systems with Volt-Var Control (VVC) being the most cost-effective solution to maintain adequate balance of voltage and power factor \cite{Souran_Safa_Moghadam_Ghasempour_Razeghi_Heravi_2016}. The use of heterogeneous equipment from several vendors to achieve a healthy balance of voltage and power within acceptable range raises several security issues which can be exploited by stealthy adversaries. As such, \cite{6859265} have proposed a white-box stealthy attack model whereby an adversary intercepts the communication between measurement devices and the central controller to inject false data measurements which successfully evades Bad data detectors such that the controller issues sub-optimal commands to the Load Tap Changer (LTC) and capacitors. Experimental results on an actual distribution model using GridLab-D revealed that VVC reduces voltage with an error rate of 2\% which is significant enough to disrupt the grid operation. The work in \cite{JuAdversarialDistributedVoltage} proposed the injection of corrupted reactive power measurements into a set of DER devices with the attacker having complete knowledge of the network topology in view of causing severe voltage mismatch within the distribution network. Simulations on a single-phase 12 kV 16-bus distribution feeder revealed that the proposed attack is not very sensitive to the step size and voltage disruptions fluctuating up to 200\% are achieved. Furthermore, the authors concluded that a topology-agnostic attack can leverage legitimate buses' responses to further boost damages. \cite{8667289}'s work  focused on the manipulation of distribution feeder voltage profiles through the injection of measurement data with gross errors into smart meters which is formulated as an optimization problem using Mixed Integer Linear Programming (MILP). Evaluations on a modified IEEE 33-bus distribution test system with one On-LTC, nine Capacitor Banks (CBs), four PV systems and 32 smart meters highlight that there is a 1\% increase in the voltage magnitude along with a 16\% increase in the On-LTC tap position which results in abnormal feeder voltage profile in both the physical and cyber layers. \cite{Shen_Liu_Xu_Lu_2021} extended the previous work by modeling a load redistribution attack on VVC as a single-leader-multi-follower bi-level mixed-integer linear programming (BMILP) model to maximize voltage profiles and increase load curtailment costs. Real-world experiments on a High-and-medium-voltage distribution system (HMVDS) in China demonstrate fluctuations on the reactive power support and voltage profiles by up to 83\%.

\subsection{State Estimation}

While state estimation has been actively used within transmission system, the transition to sustainable energy sources introduced Distribution System State Estimation (DSSE) to estimate distribution system variables in real-time with highest possible accuracy and monitor the distribution feeder operations in power grids \cite{7779155}. However, extending traditional state estimation approaches to active distribution systems poses several operational challenges such as observability problem, unbalanced operations, etc. as well as cyber security concerns \cite{8466598}. In this view, we present the state-of-the-art threats on DSSE based in the two state estimation phases namely:

\subsubsection{Balanced Single-Phase Distribution}

While distribution networks are often unbalanced and single state estimations within such systems provide sub-optimal results \cite{8282458}, it is worth noting that distribution feeders have low $x/r$ ratio which is a reason why data integrity attacks on DSSE have not much been explored \cite{8442503}. However, \cite{8307441} 
proposed a practical non-linear false data injection on voltage measurements of nodes to compromise DSSE while also enabling an adversary to infer the system state from power flow or power injection measurements without much hindrance. Simulations on a single-phased balanced IEEE 56-node test feeder showed that the approximation of system state is very close to the accurate system state with a relative error of up to 7\% for voltage magnitude and 0.6\% for voltage phase angle. Furthermore, under the attack, the state estimation is successfully brought down by nearly 6\% of its original value.

\subsubsection{Unbalanced Multi-phase Distribution}

As earlier mentioned, unbalanced multi-phase distribution (more specifically three-phase distribution) is the most optimal state estimation solution. However, from the work proposed by \cite{8307441}, it can be seen how a simple corrupted measurement attack can negatively impact the state estimation in balanced single-phase distributions. Therefore,  \cite{8625609} introduced a three-phased coupled corrupted measurement injection attack on a local state-based linear DSSE for multi-phase and unbalanced smart distribution systems which considers the weak couplings among phases to reduce the number of measurement modifications. After evaluation of the proposed attack on an IEEE 13 and IEEE 37 Bus Test Feeders shows that the DSSE under the proposed attack results in approximately similar Largest Normalized Residual (LNR) as that without attacks under 100 Monte-Carlo simulations. However, the DSSE under simple attacks proves to be highly effective by projecting a LNR of 5 which is higher than the LNR threshold. On the other hand, \cite{9035401} proposed a load redistribution attack on a closed-loop conservation reduction in an unbalanced three-phase distribution network integrated with DERs using MILP to inject malicious measurements into smart meters communication in view of rising the three-phase active power flow at the substation. Simulations on an IEEE 12-node test feeder with 1 OLTC, 2 PVs, 2 CBs and 17 smart meters revealed the OLTC tap position increases by 4 after the attack which increases the feeder voltage profile and in turn increases customer energy consumption. Furthermore, at node 3, the voltage physical layer is 0.94 which slightly violates below the minimum voltage limit and can therefore result in premature breakdown of electrical appliances.

\section{Threats on Energy Billing \& Trading}
\label{sec:energybilling}

\begin{table*}[]
\begin{tabular}{|p{2.5cm}|p{3.5cm}|p{2.5cm}|p{3cm}|p{3cm}|}
\hline
\textbf{Main Category} & \textbf{Ref No.}  &\textbf{Attack Target}  &\textbf{Attack Type} &\textbf{Attack Mechanism}\\ \hline
\textit{\textbf{\makecell{Distribution Lo-\\cational Margi-\\nal Pricing}}} & \cite{Zhang_Wang_Li_2019}  & Bus voltage measurements & Short-term Data Injection & Dinkelbach non-convex optimization  \\ \hline
\multirow{4}{*}{\textit{\textbf{\makecell{Real-time Prici-\\ng}}}} 
 & \cite{10.1145/2508859.2516705}  & Price Data Packets & Scaling/ Delay & Control-theoretic clock manipulation \\ \cline{2-5} 
 & \cite{7374725}  & Price signals & Short-term Data Injection & Stackelberg game \\ \cline{2-5} 
 & \cite{7440873}  & Smart meter data & Short-term Data Injection & Lagrangian Optimization \\ \cline{2-5} 
 & \cite{7401113}  & Price Signals & Long-term Scaling/ Delay/ Data Injection & Control-theoretic Linear Optimization \\ \hline
 
\multirow{5}{*}{\textit{\textbf{\makecell{Transactive En-\\ergy Market}}}} 
 & \cite{8440471}  & Cap price, Bid price, Demand \& Breaker Operations & Proxy Attacks & Tampering with information from controller   \\ \cline{2-5} 
 & \cite{8667445}  & Electricity Prices & MiTM Short-term Data Injection &  Simple least square approach for demand co-efficients Estimation  \\ \cline{2-5} 
 & \cite{8919711}  & Customer Bids & Fake Bidding Injection & Approximation of aggregate functions using market equilibrium information. \\ \cline{2-5} 
 & \cite{9089079}  & Customer Bids & Short-term Data Injection &  Biased welfare function maximization using market equilibrium information. \\ \cline{2-5} 
 & \cite{9274708} & Gateway between prosumers and system & Discard/Delay/ DDoS & Approximation of aggregate cost function using quadratic optimization \\ \hline
 
\multirow{2}{*}{\textit{\textbf{\makecell{P2P Distributed \\Energy Trading}}}} 
 & \cite{Islam_Mahmud_Oo_2018}  & Power generation and usage patterns & Short-term Data Injection & Constraint Optimization to minimize energy demand and sale.  \\ \cline{2-5} 
 & \cite{9248761}  & Demand & Demand Data Manipulation & Game theory \& Smart Meter Tampering \\ \hline
\end{tabular}
\caption{Comparative View of Threats on Energy Billing \& Trading.}
\label{tab:threatsbilling}
\end{table*}

Within the past two decades, traditional power systems have transitioned from a centralized supply side approach to a decentralized supply and demand side management due to power system restructuring and integration of smart distribution systems with renewable power sources \cite{abidin2018secure}. However, such increased complexity exposes energy pricing and trading to cyber threats from several adversaries for their own revenue gains or for other malicious intentions \cite{Aitzhan_Svetinovic_2018}. Therefore, we present a taxonomy of threats on the energy billing and trading process of smart grids as in Table \ref{tab:threatsbilling}. 

\subsection{Distribution Locational Marginal Pricing}

Following the success of locational marginal pricing within transmission systems \cite{9406028}, the adoption of Distribution Locational Marginal Pricing (DLMP) enables reduction in end-user energy costs, efficient peak-demand stress management on utilities and enhanced system sustainability \cite{7862921}. However, degree of accuracy of DLMP is dependent on the integrity of the Distribution System State Estimation data \cite{Zhang_Wang_Li_2019} which exposes DLMP to data integrity attacks. Zhuang and Liang \cite{Zhuang_Liang_2019} were the first to study the effects of directly injecting corrupted data of small magnitude to the bus voltage measurements of DLMP by formulating a non-convex optimization problem, which is solved by Dinkelbach’s algorithm. Experimental validations on a modified multi-phase and unbalanced IEEE 13-bus test feeder revealed that an adversary can heavily benefit from a sharp decrease in pricing (from 10.07¢/kWh to 0.76¢/kWh) at the attacked bus while the total payment of all other customers increases from 34873 cents to 35078 cents.  

\subsection{Real-time Pricing}

One of the key price response mechanisms of demand side management is Real-Time Pricing (RTP) which enables efficient power utilization and reduction in electricity costs \cite{Dai_Gao_Gao_Zhu_2017}. Due to the complex closed-loop feedback control approach which is used to maintain grid stability and performance \cite{Gusrialdi_Qu_2019} along with market expansions for efficient Demand Response programs \cite{Cioara_Anghel_Bertoncini_Salomie_Arnone_Mammina_Velivassaki_Antal_2018}, adversaries can easily affect energy markets through simple yet powerful attacks. \cite{10.1145/2508859.2516705} demonstrated the impacts of modifying the incoming price signals to a group of customers during transmission via either a reduction of values during scaling attacks or providing old prices during delay attacks. Upon validation of their proposed attacks on a normally-distributed  Constant Elasticity of Own-price (CEO) model for each customer of New South Wales, Australia's half-hourly total demand load for 2013 as baseline load, the authors concluded that under scaling attacks, excessive distribution line overload events occurred and system volatility was directly proportional to the proportion of customers under attack and inversely proportional to amplification. Furthermore, delay attacks increased the distribution line overload which causes circuit breakers to open and eventually leads to marginal system instability and regional blackouts. 

The work illustrated in \cite{7374725} focuses on arbitrarily increasing the price signals from the electric provider to the smart meters with the aim of maximizing the mismatch between energy supply and demand. Using real-world Summer 2004 Polish power flow system dataset, \cite{7374725} concluded that such attacks heavily impact load shifts and redistribution in the power grids which in turn overloads and heats up transmission lines to cause failures and catastrophic blackouts. \cite{7440873} proposed two approaches known as Ex-ante attacks and Ex-post attacks by injecting false information into demand-users or supply-users before and after the price decision making process respectively for maximizing welfare on the real-time pricing schemes. Evaluations with 15 demand-users, 20 supply-users  and a traditional power system set-up, they concluded that such cyber-threats can effectively alter real-time prices to the benefit of the adversaries. The authors in \cite{7401113} extended  \cite{10.1145/2508859.2516705}'s work by modeling a more realistic attack model whereby an adversary can compromise the integrity of price signals repeatedly over a long period of time and at any moment as opposed to being constrained to one-shot scaling or delay attacks in view of maximizing the gap between generated and consumed power. Experimental results highlight that their additive approach causes greater damage to the market stability and is more powerful while attacking sensed data by smart meters. 

\subsection{Transactive Energy Market}

\begin{figure}
  \centering
  \includegraphics[width=8cm]{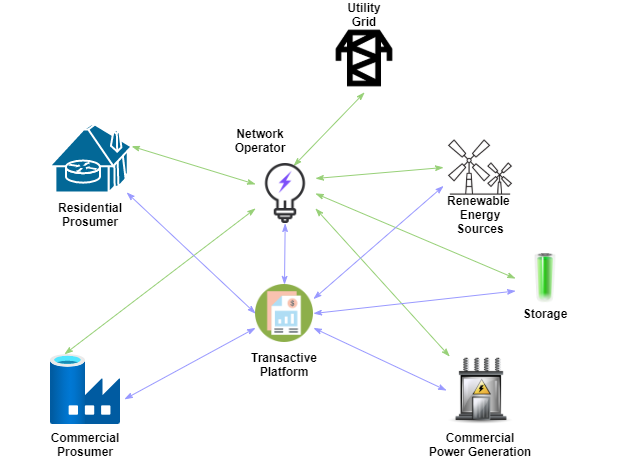}
  \caption{Conceptual Model of a Transactive Energy Market. Image adopted from \cite{8926947}. The figure represents a block architecture of a Transactive Energy Market as a collection of economic and control systems that enables the dynamic balance of supply and demand of electricity across the entire electrical infrastructure using value as a key operational parameter \cite{8940952}.}
\end{figure}

In recent years, transactive control \cite{8940952} has been extensively studied to enable the integration of DERs and Renewable Energy Sources (RES) with smart grids to ensure safety and efficient operability as it promises the flexibility of responsive assets in the grid and maintains a dynamic balance of energy supply and demand. However, experts fear that the participation of the prosumers at
the edge \cite{8738802} can lead to several security and privacy concerns due to frauds, unfair welfare maximization, etc. In this view, \cite{8440471} studied the effects of three data integrity proxy attacks namely the manipulation of cap price from 3.78\$ to 0.01\$, manipulation of bid price and quantity on total demand, and lastly, the manipulation of the breaker operations by altering router at generator substation for a set-up of thirty houses. After testing their proposed attack models on TESP framework, the authors concluded that the the manipulation of cap price as well as the alteration of bid price and quantity on total demand resulted in a spike in cooling set-point, fluctuations in the HVAC controller loads and eventually impacts overall demand on the feeder. Furthermore, the tripping of the breaker significantly affects the market price and may result in financial losses on behalf of the service provider. The work in \cite{8667445} focused changing the integrity of electricity prices as a man-in-the-middle attack during communication or as the manipulation of sensed electricity usage data at smart meters within a transactive energy market. Simulations on the IEEE 69-bus test system revealed that attacks on electricity prices have higher implications than attacks on consumption data by resulting in higher demand fluctuations and hence, higher voltage violations. More specifically, a 10\% reduction of electricity prices increases the magnitude of oscillation of energy demand and distribution system voltage. \cite{8919711} proposed the formulation of an adverse generator to manipulate the bids of other customers in the view of shifting the transactive market equilibrium to maximize welfare gains. Experimental validations on GridLAB-D and PNNL set-ups demonstrate the attack's success in increasing the adversary's monetary gains whilst also heavily impacting the social welfare of other customers if attack parameters are wrongly estimated. The authors in \cite{9089079} extended the same approach in \cite{8919711} by imposing restrictions on the adverse generator such that the adversary has to initially protect his own assets from operation states and to hide a successful attack as long as possible to maximize profit gains. After compromising 80\% of the HVAC systems, the electricity prices and total energy traded increased significantly, and the positive gains are upped with increasing attack intensity causing financial losses of other customers. However, with increasing attack intensity comes a decrease in the adversary's marginal returns. \cite{9274708} proposed three attack scenarios targeting the gateways between prosumers and the system within a transactive energy market. During the first threat model, the attacker gains access to a gateway in order to delay or discard the bids based on bidding information, the second scenario involves discarding or delaying selected bids without complete information and thirdly, the adversary launches a Distributed Denial of Service (DDoS) attack. The authors concluded that such simple attacks can effectively alter the clearing price of a blockchain-based transactive market.

\subsection{Peer-to-peer Distributed Energy Trading}

The decentralization of energy market models enables local nodes to exchange surplus power in Peer-to-Peer (P2P) setting which results in major financial welfare for both the prosumers and the utility company \cite{8513887}. However, the complexity of such large scale decentralized energy trades within untrusted and non-transparent energy markets is highly vulnerable to several security and privacy concerns \cite{8234700}. In this view, \cite{Islam_Mahmud_Oo_2018} developed an optimal false data injection attack to enable adversaries to peak at the power generation and usage patterns for extracting the maximum benefits from legitimate nodes while minimizing the gap between power sold and power purchased to avoid detection. Results after validation of the proposed attack approach on a residential micro grid with four households revealed that the proposed attack significantly impacts the profits of legitimate houses by 86\% to 94\%. The work in \cite{9248761} proposed the manipulation of prosumers' demands by a malicious supplier acting as a participating prosumer to maximize their financial welfare which can be achieved either through smart meters tampering or communication network interception. Experimental evaluations with a real dataset from Austin, Texas revealed that attacking 70\% of prosumers decreases the average utility for prosumers by 2.7\% and the profits of the external energy supplies by 10.6\%.

\section{Main Research Gaps \& Future Directions}
\label{sect:future researchgaps}

Since the topic of FDI threats within active distributions systems is an emerging topic of research with very few related studies at the time of writing, we present the main existing research gaps and provide some recommendations for fueling future research within this field.

\subsection{Existing Research Issues}

In what follows, we discuss some of the main gaps in the current FDI attack studies within active distribution systems.
\begin{enumerate}[wide, labelwidth=!, labelindent=0pt]
\item \textit{Limited FDI threats studied in perspective of active distribution systems}: While most research on FDI threats on smart power grids have concentrated mostly on transmission networks, there is currently a considerable lack of FDI attacks proposed on ADSs. The literature surveyed within this manuscript tried to cover most or if not all of the integrity attacks on distribution networks. However, there are still several open challenges with respect to the scope. For instance, at the time of writing, only \cite{9284586} assessed the impact of FDI attacks on energy storage infrastructure and SoC information. Hence, it is vital for researchers to extend the aforementioned work in the aim of exposing the subtle extreme vulnerabilities of energy storage systems. Similarly, our survey highlighted that not much studies have been undertaken which properly assess the impact of FDI threats on IEDs, $\mu$-PMUs, DLMP, etc. 

\item \textit{Lack of realistic real-world experimentation}: The FDI threats on ADSs proposed by existing literature are produced and evaluated within laboratory confined settings with several assumptions such as linearity. However, industrial standards differ from those studies such that models may be non-linear and are alternating current based systems. Therefore, we believe that more realistic FDI attacks must be formulated against large-scale realistic industrial networks/systems with lesser assumptions. 

\item \textit{Lack of corroboration of FDI attack evaluations}: Even though the FDI attacks proposed within the existing related literature has successfully proven their impacts through numerical evaluations against bench-marked test cases, there is still a lack of experimental result validations on standardized testbeds. Testbeds are vital for assessing the performance of attacks on power grids which take into consideration the architectures, security concepts, etc. Hence, we believe that before researchers plan to carry further research on this emerging topic, it is of high priority to initially set some standardization which will enable easy comparison of studies. 

\item \textit{Lack of state-of-the-art methods to detect FDI attacks}: Over the years, several techniques have been proposed to overcome the threats imposed by FDI attacks on smart grids. For instance, Kullback Leibler distance method, fast go-decomposition, unscented Kalman filter (UKF), Bayesian formulation, etc. have been previously proposed as effective countermeasures against FDI attacks \cite{ Sayghe_Hu_Zografopoulos_Liu_Dutta_Jin_Konstantinou_2020}. However, with the increasing strength and stealthiness of recent FDI attacks, several of the previously proposed solutions have become ineffective and outdated. In this view, we believe that state-of-the-art computer science techniques must be applied to enable effective detection of FDI threats. Some of the recent works in the field FDI counter defenses that have recently surfaced include the use of state-of-the-art computer science techniques such as:
\begin{itemize}[wide, labelwidth=!, labelindent=0pt]
    \item AI-based solutions: Artificial Intelligence has proven its success in several fields and disciplines and has emerged as one of the most successful technologies of the 21st century. Traditional BDDs that usually calculate the norm of the residuals between an attacked state and a normal state tend to fail in presence of stealthy FDI attacks. In order to overcome the FDI attack detection limitations posed by the use of traditional BDDs \cite{Sayghe_Hu_Zografopoulos_Liu_Dutta_Jin_Konstantinou_2020}, researchers have also developed several solutions by leveraging machine learning and deep learning techniques to effectively detect such cyber threats. For instance, authors in \cite{7063894}, which is among one of the earliest works, investigated the use of supervised machine learning algorithms for FDIA detection. The authors proposed the utilization of popular supervised and semi-supervised learning algorithms to detect FDI attacks with better performance than traditional BDDs. In response, other works \cite{8214282, 9333913, Mohammadpourfard_Sami_Seifi_2017, 8861919} have followed which employed several machine learning and deep learning techniques for similar tasks. For instance, Kumar et al. \cite{9333913} similarly employed various machine and ensemble learning mechanisms for FDI attack detection. Several milestones have been achieved in developing FDI detection mechanisms through the use of machine learning technologies. However, throughout the past decade, there has been increased interests and applications of deep learning neural networks in several domains which have shown improved accuracy rates over traditional machine learning algorithms \cite{9395437}. Indeed, researchers within the field of FDI attack detection followed in the same footsteps. The work in \cite{7684102} is one of the earliest to employ the use of deep learning networks  to analyze real-time measurements from the geographically sparse PMUs. Another work proposed by Haftu et al. \cite{9786629}  developed a deep learning model for detection of false data injection attacks using a new using a new data-driven State Estimation model. Similarly, several other works \cite{8791598, 9144530, 9110395, 8450487} have been proposed with improved deep neural networks. 
    
    \item Graph Signal Processing-based solutions: While machine learning and deep learning techniques have been studied for FDI detection, the main prevailing issue with such approaches is the need for historic datasets that also contain attacked states which is extremely hard to obtain. Therefore, to overcome such problem, recent trends in literature have highlighted the use of graph signal processing techniques to detect FDI anomalies in power grid systems. Graph signal processing is an emerging field where classical signal processing tools developed in the Euclidean domain have been generalised to irregular domains such as graphs. \cite{8784391} first proposed the use of anomaly filtering using Graph Fourier Transform to detect stealth FDI threats. Similarly, the authors in \cite{8969373} developed a countermeasure to FDI threats in synchrophasor measurements through empirical evidence that PMU information are low-pass graph signals and the use of the features of the PMU graph signal. Similarly, \cite{9582826}'s study utilized a scalable Graph Neural Network (GNN) for FDI attack detection. While most of the work is based on transmission systems where the graphs are connected in nature, it would be great to investigate the extension of the work more from a distribution system perspective where the topology is radial or tree like. As very few studies have been undertaken within this area, we believe that this is a very interesting research gap that can be further explored.
    
    \item Hybrid solutions: Since real-time smart grid systems have become highly susceptible to real-time FDI attacks, researchers have began implementing hybrid countermeasures by leveraging hybrid solutions that enhances the chance of FDI attack detection. Data-driven solutions tend to leverage spatio-temporal correlations among multi-time-instant sensor measurements to detect FDI outliers. A few works including \cite{ 9706368, 9352502, Roy_Debbarma_Guerrero_2022} on data-driven FDI attack detection haven been proposed in literature. For instance, the work in \cite{9706368} design an unsupervised detection scheme to detect the stealthy attack by employing a deep auto-encoding Gaussian mixture model which also means the data imbalance tolerance. While data-driven-based FDI defense mechanisms have been studied in the recent past, a new hybrid approach, namely physics-based solutions, has surfaced amongst researchers within this topic. \cite{Jevtic_Zhang_Li_Ilic_2018} is a pioneering work that uses a hybrid physics-based deep learning approach for FDI threat detection in power systems. In similar line, \cite{8974027, 8945588, https://doi.org/10.48550/arxiv.2204.12970} are among the few works that employed the use of physics-based methods for FDI attack detection. 
    
\end{itemize}

\item \textit{Lack of investigation of FDI threats in black-box settings}: Traditionally, the crafting of FDI threats by an adversary requires complete knowledge of the system. However, in practice, it is extremely hard to obtain topologies and other valuable characteristics of power systems as they are usually secured and encrypted to prevent unauthorized access. As prior knowledge for crafting FDI attacks is difficult to obtain in real-life scenarios, researchers have been motivated to contribute towards data-driven blind attack strategies based on black-box modelling for smart grids. In that regard, some few works \cite{6503599, 8323244, 8425789, 7741928, 8581440, 9506908} have demonstrated that even with limited amount of information, successful FDI attacks have been implemented on transmission systems. Furthermore, the works in \cite{7741557, Anwar_Mahmood_Tari_Kalam_2022, Anwar_Mahmood_Pickering_2016} contributed significantly towards the development of blind FDI attacks in black-box based settings. Specifically, in a black-box setting as shown in Figure \ref{fig:whitevsblack}, an attacker is able to construct stealthy attack vectors which are based on the measurement subspace of the sensor information, thus eliminating the need of prior "known" system information. However, the mentioned studies have mostly been undertaken on transmission systems. Therefore, due to the major differences in characteristics between power transmission and distribution systems, transmission-based FDI attack construction mechanisms may not be extended to distribution systems. Very limited studies including \cite{9632316} have been undertaken to investigate the crafting of black-box FDI attacks in distribution systems. Thus, we believe that there exists a very critical gap in the existing state-of-the-art to be addressed by researchers as in the real world, adversaries often need to craft black-box attacks as they lack valuable insights of a particular system.

\begin{figure*}
    \centering
    \subfloat[\centering Demonstration of an FDI attack with known prior information. (An adversary constructs a FDI attack using known system information which produces a residual norm similar to a normal behaviour scenario)]{{\includegraphics[width=8cm]{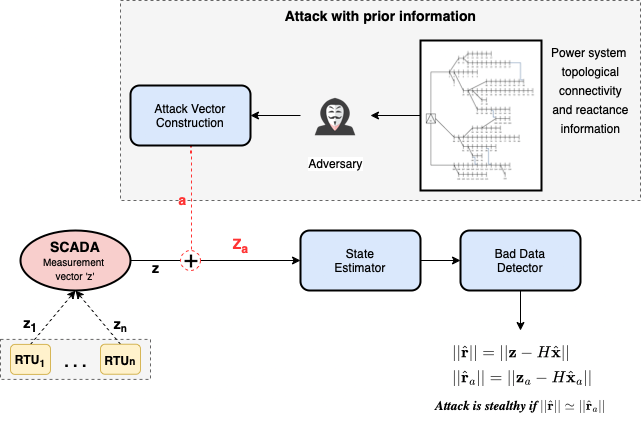} }}%
    \qquad
    \subfloat[\centering Demonstration of an FDI attack without prior information. (An adversary constructs a blind FDI attack using subspace information of the measurement data.]{{\includegraphics[width=8cm]{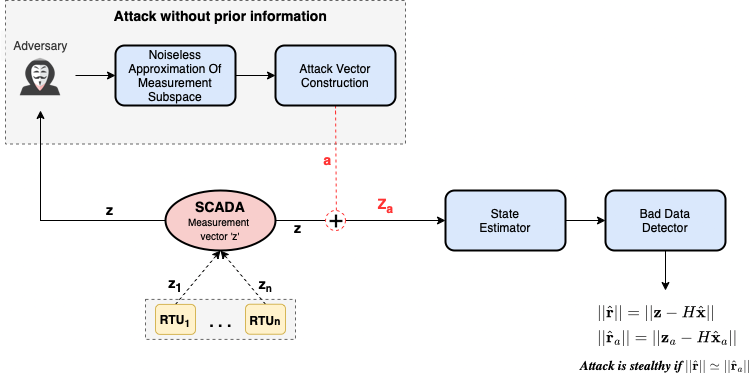} }}%
    \caption{Comparison between FDI attack construction with prior information vs. in a black-box setting (referred from \cite{Anwar_2017}). \label{fig:whitevsblack}}%
\end{figure*}

\end{enumerate}

\subsection{Future Directions}

With the number of cyber-threats constantly rising on smart grids, securing modern active distribution networks is becoming one of the top agendas of several nations. Within this section, we recommend some future research prospects in relevance to FDI threats on ADSs.

\begin{enumerate}[wide, labelwidth=!, labelindent=0pt]
    \item \textit{Secure Communication \& Aggregation Protocols}: Active distribution networks feature bi-directional flow of critical data and messages. Throughout this review, we have uncovered that AMI communications can easily be subverted to inject corrupted measurements into the data \cite{Siqueira_Ramos_Neves_Canha_2018}. Furthermore, attackers are enable to actively participate in the data aggregation process to input falsified data into the network \cite{Wang_Lu_2013}. The critical issue to be addressed is how to accurately identify subtle false data injection attacks and refrain adversaries from maliciously gaining access to the network. How to design strong data encryption and differentially private schemes with a healthy trade-off between data utility and accuracy along the aggregation path is also a challenging issue in smart grids.
    
    \item \textit{Privacy Preservation Consideration}: Most, if not all, of the FDI threats proposed within the emerging field of active distribution systems are mainly concerned about compromising the stable operations of grids. However, we believe that researchers should actively research on privacy breaching attacks and their countermeasures to expose and resolve the vulnerabilities of future active distribution networks. One such solution is the application of collaborative learning within the power distribution infrastructure.
    
    \item \textit{FDI attacks on Blockchain-based distribution system solutions}: Since the rise of blockchain paradigm, energy system researchers have been actively finding blockchain-based solutions for smart grids, more specifically in the field of distribution systems. While blockchain offers several promising benefits of security and complex interaction modeling, FDI attacks on blockchain based solutions for distribution networks must be thoroughly studied as it is itself a very new research field.
    
    \item \textit{FDI threats on Distributed Energy Management}: DERs are changing the way of producing and managing electricity in several countries. Rather than the generation of electricity by centralized power stations, it is currently being generated through renewable energy units or systems that are commonly located at homes or businesses. Distributed Energy Management is an emerging topic within the grid research community and therefore, it is vital to study the emerging FDI risks in its relation.
    
    \item \textit{Lightweight security mechanisms}: Within the smart grid ecosystem, there are several resource-limited sensor devices (e.g. smart meters, etc.) which consistently gather valuable data. Therefore, due to the lack of computational resources, conventional security mechanisms are not suitable for sensor nodes in the smart grid system. Therefore, to defend against FDI threats, researchers must focus on developing light-weight defense solutions to overcome the resource-limitation of sensor devices.
    
\end{enumerate}

\section{Conclusion}
\label{sect:conclusion}

Active distribution systems in smart grids are being threatened from an emerging class of cyber attacks known as False Data Injection attacks. Through the injection of corrupted measurement vectors, attackers can easily bypass bad data detection countermeasures and compromise the availability, integrity and confidentiality of the data within ADSs. Moreover, properly coordinated and executed FDI cyberattacks can have devastating impacts on not only the distribution system, but on the overall smart grid due to large-scale power system operation failures, regional blackouts, energy thefts, and so on.

Therefore, in this manuscript, we presented a survey of of FDI attacks on active distribution systems and proposed a taxonomy to classify the studies with respect to four attack targets namely end-user level, field devices, control center and, energy pricing and trading. Finally, we identified some main gaps in the existing research and provided some future research directions for FDI attacks on active distribution systems.

\bibliographystyle{cas-model2-names}
\bibliography{cas-refs}
\end{document}